\documentclass[%
 reprint,
 amsmath,amssymb,
 aps,pra
]{revtex4-1}

\usepackage{graphicx}
\usepackage{dcolumn}
\usepackage{bm}

\begin{document}

\title{Frequency measurement of the clock transition of an indium ion sympathetically-cooled in a linear trap}

\author{Nozomi Ohtsubo, Ying Li, Kensuke Matsubara, Tetsuya Ido}
\author{Kazuhiro Hayasaka}%
 \email{hayasaka@nict.go.jp}
\affiliation{%
National Institute of Information and Communications Technology (NICT),\\
4-2-1 Nukuikitamachi, Koganei, Tokyo 184-8795, Japan
}%

\date{\today}

\begin{abstract}
We report frequency measurement of the clock transition in an $^{115}$In$^+$ ion sympathetically-cooled with Ca$^+$ ions in a linear rf trap. The Ca$^+$ ions are used as a probe of the external electromagnetic field and as the coolant for preparing the cold In$^+$. The frequency is determined to be 1 267 402 452 901 049.9 (6.9) Hz by averaging 36 measurements using an optical frequency comb referenced to the frequency standards located in the same site.
\end{abstract}

\maketitle

\section{Introduction}
Recent progress of optical clocks based on neutral atoms in optical lattices and on single ions in rf traps has led to fractional uncertainties of 10$^{-18}$ level\cite{RMP2015}.
Various implementation with different atomic species are being explored to reach and/or to reduce these uncertainties.
It is pointed out that measurement of frequency ratio of different optical clocks is an essential step towards the redefinition of the second\cite{Hong2017}.
The ratio measurement is of crucial importance also for testing the temporal variation of the fundamental physical constants\cite{RMP2015}. 
In the implementation with single ions, the optical frequencies have been reported so far  with $^{171}$Yb$^+$\cite{Huntemann2016,YbNPL}, $^{88}$Sr$^+$\cite{SrNRC,SrNPL},  $^{40}$Ca$^+$\cite{CaInnsbruck,CaNICT,CaWuhan}, $^{199}$Hg$^+$\cite{HgNIST}, $^{27}$Al$^+$\cite{AlNIST},  $^{115}$In$^+$\cite{OL2000,OC2007}. 
In$^+$ is assumed to be a promising candidate of the original single-ion optical clock proposed by Dehmelt\cite{Dehmelt1982}, and frequency measurements with fractional uncertainties in 10$^{-14}$ level have been reported previously\cite{OL2000,OC2007}.
However, there has been no update since 2007 in contrast to other ions reporting repeated improvements in the level below 10$^{-16}$.
Frequency discrepancy of about 1~kHz by two research groups remains also unresolved\cite{OL2000,OC2007}.
Recent theoretical prediction of small black body radiation (BBR) shift revives the advantage of In$^+$ as an ion of the IIIA group\cite{Sofronova11}. 
Frequency measurement of In$^+$ with more precision than those of previous reports will make In$^+$ a useful candidate of the frequency ratio measurement. 
It should be also noted that In$^+$ is a promising candidate of the multi-ion optical clock with enhanced stability \cite{Mehlstaeubler2014}.  
In this paper we demonstrate spectroscopy of the In$^+$ clock transition in a new approach based on sympathetic cooling, and report its frequency measurement using a local frequency reference, UTC(NICT) (realization of Universal Coordinated Time at NICT) and a Sr optical lattice clock. 

\begin{figure}[h]
\centering\includegraphics{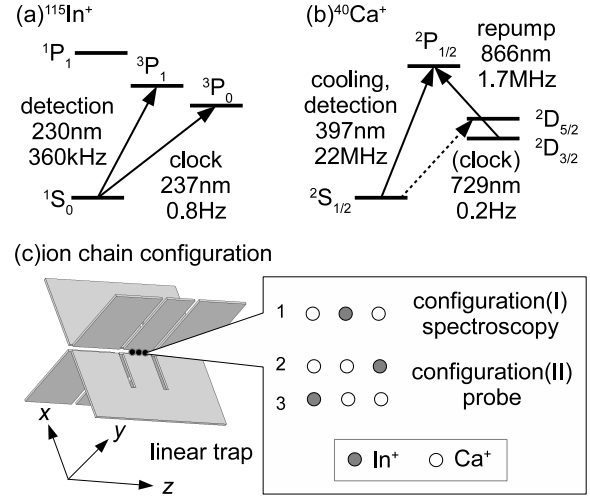}
\caption{Concept of the new implementation of the In$^+$ optical clock. (a) Relevant energy levels and transitions in In$^+$, and (b) those of Ca$^+$. (c) Schematic drawing of the ion chain configurations for our new approach.}
\label{concept}
\end{figure}

The ion-based optical clock requires methods of cooling and of state detection in addition to the access to the clock transition.
The $^1$S$_0$-$^1$P$_1$ transitions to provide the two methods in the IIIA group ions are located in the vacuum ultraviolet range, and technical difficulties to prepare the single-mode light sources prevent the implementation of the optical clocks. 
This difficulty has been resolved by quantum logic spectroscopy (QLS) of Al$^+$, in which another ion is used for the cooling and the detection\cite{AlQLS}.
The disadvantage of the QLS is high technical demands for the cooling to the ground state of motion, and repeated detection for fidelity state readout.
In$^+$ has an exceptional feature as an IIIA group ion that the $^1$S$_0$-$^3$P$_1$ transition can supply direct state detection and laser cooling owing to its relatively large linewidth of 360 kHz as shown in Fig.\ref{concept}(a).
This feature was employed to direct quantum state detection\cite{Peik1994}, and to bichromatic laser cooling\cite{Peik1999}. 
The bichromatic laser cooling requires two laser frequencies to stabilize the cooling.
The previous implementations of the In$^+$ optical clock rely on these methods\cite{OL2000,OC2007}.
We demonstrate a new approach, which combines the concepts of sympathetic cooling used in QLS, and of the direct state detection in the previous In$^+$ optical clocks\cite{Hayasaka2012}.
Choice of Ca$^+$ as shown in Fig. \ref{concept}(b) provides an easier access to cooling using compact diode-laser-based systems \cite{Hayasaka2011,Uetake2009}.
When Doppler cooling is applied to a Ca$^+$, a sympathetically cooled In$^+$ can be prepared in a simpler way than by bichromatic cooling.
The direct quantum state detection of the In$^+$ allows an easier implementation of the optical clock than by QLS.
Although the frequency inaccuracy achieved ultimately with this approach is limited by the Doppler cooling limit temperature of Ca$^+$, the sideband cooling to the ground state using the $^2$S$_{1/2}$-$^2$D$_{5/2}$ clock transition could be used to reduced the inaccuracy by time dilation\cite{Tanaka2015}. 
The Ca$^+$ can be used also as a probe for the trapping and surrounding fields.
The use of the $^2$S$_{1/2}$-$^2$P$_{1/2}$ transition with a larger scattering rate provides a fast evaluation of micromotion and the magnetic field.
More precise field evaluation might be implemented by using the clock transition.
The minimum configuration for the sympathetic cooling is to add one Ca$^+$. 
However, asymmetric application of the trapping field is necessary to locate the In$^+$ at the same site in this configuration .
It is essential to place the ion at the same site to avoid location-dependent systematic shifts.
Use of two Ca$^+$ ions instead of one enables a symmetric configuration depicted as (I) in Fig.\ref{concept}(c).
This configuration is very convenient for the clock transition spectroscopy, because it is relatively easily recovered whenever it changes into the configuration (II).
The recovery was demonstrated with amplitude modulation on the cooling laser to destabilize the configuration (II) only\cite{Hayasaka2012}.
In the present study we use an even simpler method to reduce the trapping rf voltage until the heavier ion comes back to the center. 
The configuration (II) with a Ca$^+$ at the center might be used for probing the electromagnetic field at the center for more precise systematic shift evaluation in future.
For these practical advantages we use an ion chain consisting of one In$^+$ and two Ca$^+$. 
We demonstrate spectroscopy of the clock transition using such an ion chain for the first time.

\section{The experimental setup}
\begin{figure}
\centering\includegraphics{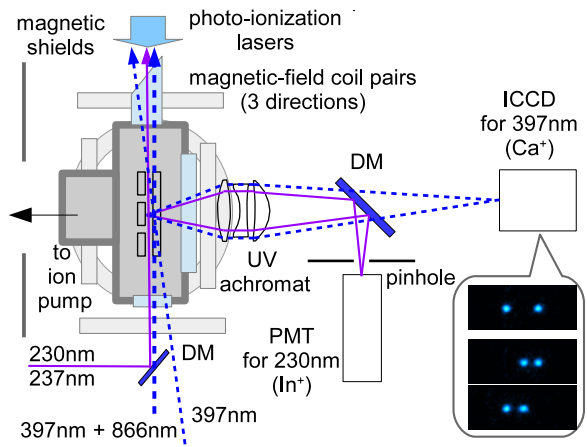}
\caption{Schematic of the experimental setup. DM: dichroic mirror, PMT: photomultiplier tube, ICCD: image-intensified CCD camera. }
\label{setup}
\end{figure}

The overview of the experimental setup is shown in Fig. \ref{setup}.

\subsection{Linear trap}
A linear rf trap is fabricated with BeCu plates with gold plating on both surfaces to avoid charging due to the deep UV beams. 
Two DC electrodes are segmented in three parts to supply confinement along the trap axis, while two rf electrodes have short slits as shown in Fig. \ref{concept}(c).
The distance from the trap center to the electrode edge is 0.5~mm, and  the width of the center DC electrode is 3~mm. 
The trap electrodes are fixed to insulators made of glass ceramic, and then attached to a support consisting of machined BeCu blocks installed in a octagon-shaped vacuum chamber made of non-magnetic stainless.  
The chamber with UV-grade sapphire windows is evacuated to a pressure below 10$^{-8}$~Pa.
The trap is driven with an rf frequency of 27~MHz and a DC with 500~V, which gives Ca$^+$ secular frequency of 4.0~MHz and 0.5~MHz for the radial and the axial directions respectively.
These voltages give radial secular frequency of 1.4~MHz for In$^+$ and in-phase mode frequency along the axial direction of 0.4 MHz for the configuration (I), which bring the In$^+$ in the Lamb-Dicke regime. 

\subsection{Light sources}
The light source system for Ca$^+$ consists of a 397-nm and a 866-nm extended-cavity diode lasers (ECDLs) for laser cooling, and of a 423-nm ECDL, and a 380-nm diode laser for resonant photo ionization.
Spectral filtering with a filter cavity or a grating, and frequency stabilization using transfer cavity referenced to a frequency-stabilized He-Ne laser are implemented to the 397-nm and 866-nm lasers \cite{Hayasaka2011,Uetake2009}.    
The light source system for In$^+$ consists of a 411-nm ECDL for resonant photo-ionizaiton, a 230-nm source for detection and a 237-nm clock laser.
The 230-nm radiation is generated by two-stage frequency doubling of an amplified 922-nm ECDL. 
The master laser is locked to a ULE (Ultra Low Expansion) cavity with a finesse of about 10,000.
About 200~mW of output from the amplifier is frequency-doubled twice, first in a periodically-poled lithium niobate (PPLN) waveguide to 60~mW at 461~nm, then in a 
beta-barium borate (BBO) crystal placed in an enhancement cavity, to the final output of 1~mW at 230~nm.  
The 237~nm radiation for the clock transition is generated by two-stage frequency doubling of an amplified diode laser at 946~nm in a similar way, except the bulk PPLN crystal in an enhancement cavity in the first stage instead of a waveguide. 
Available output power at 237~nm is 15~mW.
We stabilize the fundamental frequency of 946~nm to a high finesses ULE cavity (finesse 320,000), and control the ULE cavity at the measured zero-crossing temperature of 32.8~$^\circ$C.
The linear drift of the cavity resonance is substantially eliminated by an AOM, to which counter-drifting frequency is introduced from a DDS-based oscillator. This drift compensation  suppressed the long-term drift below  0.05~Hz/s.  

\subsection{Detection system}
Simultaneous detection of the two ion species is performed with an image-intensified CCD (ICCD) camera for Ca$^+$ and  a solar-blind photo-multiplier tube (PMT) with a quantum efficiency of 30\% for In$^+$.
The separate imaging of two wavelengths resolves the large chromatic aberration for 397nm and 230nm of the imaging lens with focal length of 50~mm and an effective aperture of 27~mm.
The configuration of the ions is continuously monitored with the ICCD camera, and is kept to the configuration~(I).
Whenever the chain changes to the configuration~(II), the measurement is interrupted and the rf potential is lowered and then heightened for reordering the ions to (I).
This simple method of ion reordering relies on mass dependence of the pseudo potential, and is effective only when the heavier ion is placed to the center.
Details of the method will be reported elsewhere.
Spectroscopic data with the configuration (II) are not recorded in the present experiments.

\section{Preparation of a cooled In$^+$}
The experimental procedure starts with optimization of the micromotion compensation voltages and the magnetic field using a single Ca$^+$ followed by loading of another Ca$^+$ and an In$^+$, then the spectra are recorded by scanning the clock laser frequency by acousto-optic modulator (AOM). 
The frequency of the clock laser is measured using an optical frequency comb during the spectrum measurement, and the clock transition frequency is determined after each measurement session by taking the average of the scan data.
The stabilization of the clock laser to the transition was not yet  implemented in the present study as the spectroscopy signal often falls unavailable due to the recurrent changes of ion configuration.   
 
The Ca$^+$ is loaded by resonant photo-ionization of thermal Ca beam from an oven installed beneath the trap using the 423-nm and 380-nm lasers\cite{Hayasaka2012}.
The ion is laser-cooled with the 397-nm and 866-nm ECDLs. 
The number of Ca$^+$ is adjusted to one for the optimization of the micromotion compensation voltages and the magnetic field.
Micromotion compensation for three directions is performed with the Ca$^+$ using the standard procedure based on the fluorescence-rf correlation technique \cite{Berkeland1998}.
The magnetic field at the trap center is controlled by three pairs of coils orthogonally attached to the vacuum chamber. 
The coil currents to null the magnetic field are searched for by using the dark states formed in the $^2$D$_{3/2}$ state of Ca$^+$\cite{Berkeland2002}. 
By iterative minimization of the Ca$^+$ fluorescence with currents of each coil pair, a set of currents to minimize the magnetic field is found\cite{Barwood1998}.
After the optimization of the fields, two Ca$^+$ are loaded in the similar way.
The number of Ca$^+$ is adjusted to two by controlling the trap rf voltage\cite{Hayasaka2012}.
Then an In$^+$ is loaded by resonant photo-ionization with the 411-nm ECDL.
The In$^+$ is sympathetically cooled with the Ca$^+$.
The number of In$^+$ is identified as dark sites in the ICCD image, and is adjusted by ejecting extra In$^+$ ions by controlling the trap rf voltage\cite{Hayasaka2012}.
By measuring the in-phase mode frequency of the chain, the final dark site is identified as In$^+$ with the mass number of 115.
The quantum state detection of the In$^+$ is performed using the PMT. 
Only when the In$^+$ is in the $^1$S$_0$ state the ion emits photons at 230~nm and gives detection of about 15 photons per 60~ms gate time with approximately 10$\mu$W of the 230-nm light.
This count clearly distinguishes that from scattered light and the dark count of 3, with a 
discrimination fidelity of about 90\%. 

\section{Spectroscopy of the clock transition}

\begin{figure}
\centering\includegraphics{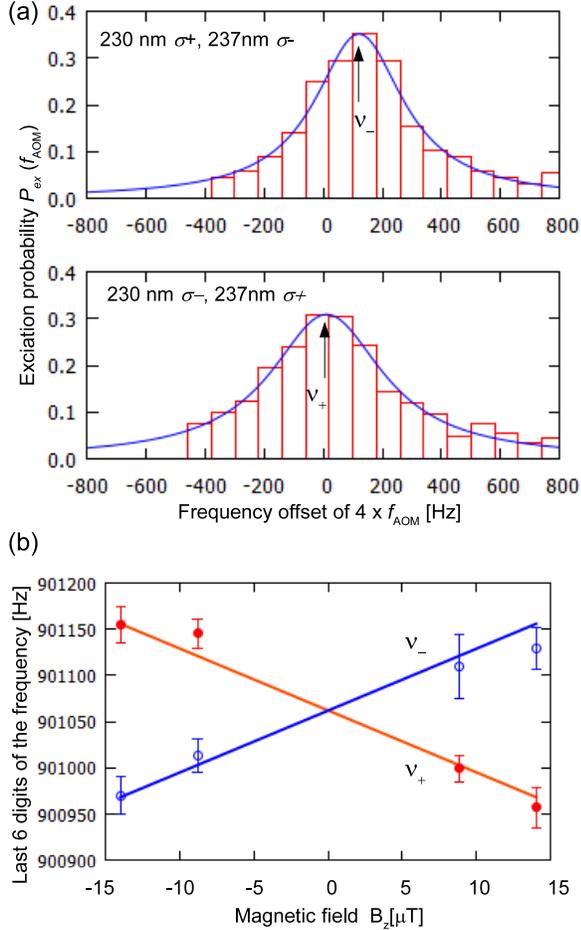}
\caption{(a) Plots of the In$^+$ clock transition spectra for the two settings of the laser polarizations. The bars represent measured excitation probability, while the curves are Lorenz curves for the least square fitting. (b) Observed dependence of the spectrum center on the magnetic field along z-direction $B_z$. The red filled circles shows those for $\sigma-$ clock excitation, while the blue empty circles represents those with  $\sigma+$ clock excitation.  The lines are linear fit to the data.}
\label{f-spectra}
\end{figure}

Spectroscopy of the clock transition is performed by recording the excitation probability as a function of the AOM frequency ($f_{AOM}$) for scanning the clock laser frequency.
In order to avoid shifts due to the lasers other than the clock laser, the Ca$^+$ cooling lasers as well as the 230-nm In$^+$ detection beam are blocked by mechanical shutters during the clock pulse irradiation periods.
200 trials are made for a single $f_{AOM}$ to obtain the excitation probability in the following procedure.
The first step is to prepare the In$^+$ in the $^1$S$_0$ state in the configuration (I).
This is confirmed by the image on the ICCD and the count of the PMT.
Then all laser beams are blocked by the shutters, and the 237-nm clock pulse for a duration of 20~ms illuminates the In$^+$.
The clock laser power of about 1.1~$\mu$W before the shutter and the estimated spot size diameter 100~$\mu$m at the trap center gives similar intensity as in the previous reports\cite{OC2007}.
After the clock pulse irradiation other lasers are turned on, and the quantum state of the In$^+$ is determined by the 230-nm fluorescence for the detection period of 60~ms.
The single cycle takes 100~ms on average including dead times of 0.8~ms to wait for the mechanical shutters to be settled and process time of about 20~ms by PC. 
While the shutter timings are precisely generated by a delay generator, the process time by PC fluctuates due to the non-real-time nature of the operating system. 
Repeating this procedure typically 200 times the excitation probability for a single AOM frequency $P_{ex}(f_{AOM})$ is obtained. 
By scanning $f_{AOM}$ with a step of 80~Hz at 237~nm the spectrum of the In$^+$ clock transition is obtained.
A $\lambda$/4 plate attached to a motorized rotation mount switches the polarizations of 230~nm and 237~nm between the settings of ($\sigma$+, $\sigma$-) and ($\sigma$-, $\sigma$+).
The spectrum of the transition from $| ^{1}S_{0}, m_{F}=9/2 \rangle$ to $| ^{3}P_{0}, m_{F}=7/2 \rangle$ is first recorded in the former setting, then that from $| ^{1}S_{0}, m_{F}=-9/2 \rangle$ to $| ^{3}P_{0}, m_{F}=-7/2 \rangle$ in the latter setting.
The transition centers ($\nu_{-},\nu_{+}$) are derived by curve fittings to Lorentzian.
The average of $\nu_{-}$ and $\nu_{+}$ gives the clock transition frequency $\nu_{0}$ with the 1st order Zeeman shift canceled \cite{SrNPL,CaNICT}.
In the following the set of the two records is defined as one measurement. 
A snapshot of the measurement with a magnetic field along $z$-axis ($B_z$) of about 8.8~$\mu$T is shown in Fig. \ref{f-spectra} with Lorenz curves for the least-square fitting.
The linewidth obtained by the fitting ranges from 350 to 700~Hz.  
We confirmed that Zeeman-shift coefficient of $\nu_{-}$ and $\nu_{+}$ agrees well with the reported values of the two transitions \cite{OL2000,OC2007}. 

\section{Frequency measurement}
\begin{figure}
\centering\includegraphics{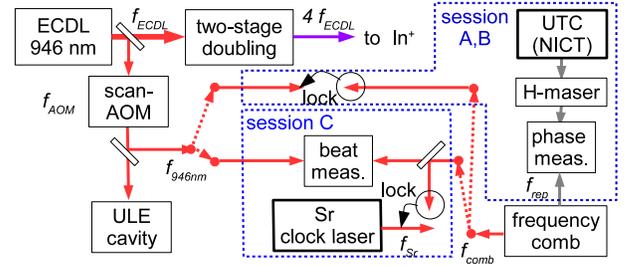}
\caption{Schematic of the clcok-laser frequency measurement. ECDL: External-Cavity Diode Laser, ULE: Ultra Low Expansion, UTC(NICT): realization of Universal Coordinated Time at NICT.}
\label{f-scheme}
\end{figure}

\begin{figure}
\centering\includegraphics{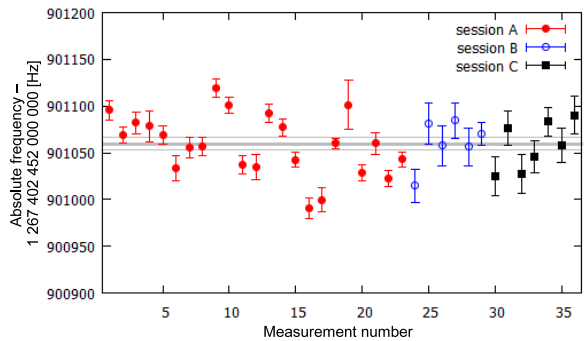}
\caption{Plot of the last six digits of the measured frequency $\nu_0$ in three sessions and of their weighed averaged value denoted as the gray line. The two gray dotted lines represent 
a range of the standard error.
Systematic shifts including gravitational shift are not taken into account in this figure. }
\label{f-frequency}
\end{figure}

Frequency determination and uncertainty evaluation have been performed as follows using the data taken in three measurement sessions, which include 23, 6, 7 measurements, respectively.
During all the sessions the clock laser frequency at 946~nm was measured with an Er-fiber-based optical frequency comb in the methods shown in Fig. \ref{f-scheme}.
The $\nu_{0}$ values of all measurements determined by the frequency comb are plotted in Fig. \ref{f-frequency}, in which error bars correspond one standard deviation of the fitting.
The mean value and the statistical uncertainty of each session are determined by the $\nu_{0}$ values by using the fitting errors as the weight. 

The comb was referenced to UTC(NICT) in session A and B, while it was referenced to a $^{87}$Sr optical lattice clock in session C \cite{Hachisu2015,APB2017}.
For the measurements in session A and B the frequency of one component of the frequency comb ($f_{comb}$) was phase-locked to the the clock laser at 946~nm ($f_{946nm}$), and we recorded the phase of the down-converted microwave relative to UTC(NICT) in every second, by which we obtained the mean frequency of $f_{946nm}$ of one measurement.
UTC(NICT) is calibrated with reference to the SI second by so-called TAI-link (International Atomic Time-link).
The fractional uncertainty of the link between UTC(NICT) and the SI second amounts to $2.6 \times 10^{-15}$ considering the uncertainty published in Circular T as well as dead time uncertainty of the UTC(NICT) and UTC-SI second\cite{Hachisu2015}. This corresponds to 3.3~ Hz for the ${\rm In}^+$ clock transition at 237~nm.  
In session C, $f_{comb}$ was phase-locked to the clock laser of the Sr optical lattice clock ($f_{Sr}$), and the beat frequency between the 946nm clock laser and the nearest comb component was recorded by a conventional frequency counter (Agilent 53132A). 
The absolute frequency of the Sr lattice clock was recently evaluated with fractional uncertainty of $9.5\times 10^{-16}$\cite{APB2017,OESubSr}. 
In session A, $B_z$ was swept from -14~$\mu$T to 
14~$\mu$T in order to confirm that the transitions of proper Zeeman sublevels were excited, and to evaluate the uncertainty of the residual 1st order Zeeman shift.
Measurements with residual magnetic field smaller than 5$\mu$~T were performed in other sessions.
The last 6 digits of the averaged frequency of each session are summarized in Table. \ref{result} with errors including statistical error and the uncertainty of the link to the SI second. 
Tiny difference of the values in session A and B shows that the simple cancellation method of the 1st order Zeeman shift is effective to this degree.
The agreement between session B and C implies that no bias is present in the frequency link to the SI second\cite{CaNICT}.
The weighted average of the sessions gives 1~267~402~452~901~060.4 (5.0), which doesn't include systematic uncertainties of the shift in the In$^+$ transition associated with fields and motion.
In general, the statistical uncertainty can be reduced by increasing the number of measurements.
However, we found that the further reduction of the statistical uncertainty by more measurement sessions would be ineffective, because the systematic uncertainty discussed in the following limits the overall frequency uncertainty in the present study.  

\begin{table}
\caption{Measurement result of the three measurement sessions. The error includes statistical and uncertainty of the link to the SI second. Systematic errors of the ion are not included. \textit{N}: number of measurements, $|B|$: magnetic field strength.} 
\label{t1}
\begin{center}
\begin{tabular}{ccccc}
\hline
Session & \textit{N}  &  $|B|$ & Reference & Last 6 digits [Hz] \\
\hline
A & 23 &  up to 14~$\mu$~T & UTC(NICT) & 901058 (8)\\
B &  6 &   $<$ 5~$\mu$~T  & UTC(NICT)  & 901065 (11) \\
C &  7 &   $<$ 5~$\mu$~T & Sr clock  & 901061 (11) \\
\hline
total & 36 &    &    & 901060.4 (5.0)\\
\hline 
\end{tabular}
\end{center}
\label{result}
\end{table}

\begin{table}
\caption{Evaluation of systematic errors. The shifts with estimated absolute value smaller than 0.1 Hz is expressed as 0 in the table. S.M.;secular motion, M.M.;micromotion.} 
\begin{center}
\begin{tabular}{lcc}
\hline
Effects & Shift (Hz) & Uncertainty (Hz) \\
\hline
1st order Zeeman shift & 0 &  4.6\\
Quad. Zeeman shift &  0 & $<$ 0.1 \\
Clock laser Stark shift & 0 & 0.1 \\
BBR shift & 0 &  $<$ 0.1\\
2nd order Doppler (S.M.) & 0 &  0.7\\
2nd order Doppler (M.M.) & 0 & 0.4\\
Gravitational shift & 10.5 & $<$ 0.1 \\
\hline
total & 10.5 &  4.7       \\
\hline 
\end{tabular}
\end{center}
\label{systematics}
\end{table}

Table. \ref{systematics} summarizes the shifts and uncertainties of the In$^+$ transition. 
The largest systematic error in the present study comes from the 1st order Zeeman shift.
Imperfect polarization of the 230~nm and 237~nm beams, and residual magnetic field along \textit{x} and \textit{y} directions allow excitation of transitions between unwanted Zeeman sublevels.
This gives asymmetry to the two spectrum centers  $\nu_{-}$, $\nu_{+}$ defining one measurement, leading to a frequency shift.  
This shift was estimated by checking the dependence of calculated center frequency $\nu_{0}$ on the magnetic field $B_z$\cite{JILA}.
From the data in session A we estimate the residual shift dependent linearly on $B_z$ with -912~Hz/mT.
The shift should be canceled at the zero magnetic field, but the measurements at low magnetic fields in session B, C show spectrum linewidths larger than 350~Hz, which suggests that possible residual magnetic field in \textit{x}- and \textit{y}-directions is not smaller than 5~$\mu$T. 
We assigned $\pm$4.6~Hz calculated for the residual field of 5~$\mu$T as the uncertainty for the 1st order Zeeman shift. 
The second-order Zeeman shift of In$^+$ is known to be much smaller than those of other ions, and is given as 4.1$\times B^2$ Hz/mT$^2$\cite{RMP2015, Herschbach2012}, in which the square of the magnetic field $B^2$ includes contributions from the residual magnetic field and the alternating magnetic field due to unbalanced currents of the ion trap rf field . 
Both contributions are estimated as much smaller than 0.1~Hz.
The AC-Stark shift due to the clock laser beam is estimated by calculating the shifts of the $^1$S$_0$ and the $^3$P$_0$ states due to coupling to the energy levels listed in ref. \cite{NISTASDver5}.
We added uncertainty of 0.1 Hz as the upper limit obtained by the calculation.
The contribution of the other lasers are not taken into account, because they are completely shut off during the clock laser irradiation periods by mechanical shutters. 
The BBR shift of In$^+$ is recognized as smaller than those of other optical clock candidates except Al$^+$, and is estimated as 17~mHz at 300~K\cite{RMP2015,Sofronova11,Herschbach2012}.
The largest contribution in the temperature uncertainty comes from the rf electrodes continuously exposed to the rf drive.
If we assume the temperature of the electrodes rises to as high as 200 $^\circ$C, and the In$^+$ ion sees the electrodes with 4$\pi$ solid angle, the BBR shift would be 0.1~Hz.
Considering the actual solid angle much smaller and the electrode temperature not higher than this assumption, we estimate the BBR shift uncertainty smaller than 0.1~Hz. 
The frequency shift due to the time dilation comes from the residual micromotion and secular motion of the In$^+$ ion.
The uncertainty by micromotion was estimated from the rf correlation signal using the single Ca$^+$ prior and after the frequency measurements in the standard method\cite{Berkeland1998}, and is estimated to be 0.4~Hz.
The rather large uncertainty comes from the low sensitivity of our rf correlation measurement of the radial direction ($x-y$ in Fig.\ref{concept}(c)) limited by small angle of one of the 397-nm laser beams, which could be improved in future by modifying the beam irradiation geometry.
Strict evaluation of the uncertainty due to residual secular motion by sideband spectra was not performed in the present study due to unavailability of a vacuum chamber port to probe the secular motion of In$^+$ along all the directions.  
Instead we estimated the worst case, in which vibrational wavepacket size of the Ca$^+$ is as large as the image on the ICCD camera.
The 3$\mu$m rms size corresponds to the maximum rms velocity 5.7~m/s of In$^+$ assuming the in-phase mode of the ion chain\cite{Hayasaka2012}.
This leads to the uncertainty of 0.7~Hz.
The gravitational shift is well characterized in other experiments done in the same floor of the laboratory building \cite{CaNICT, Hachisu2015}, and is 10.5~Hz for In$^+$.
The systematic shifts associated with the clock frequency scan is evaluated to be sufficiently small compared to other shifts. 
The clock transition frequency is determined to be 1~267~402~452~901~049.9~(6.9)~Hz from the measurements and the above discussions.

\begin{figure}
\centering\includegraphics{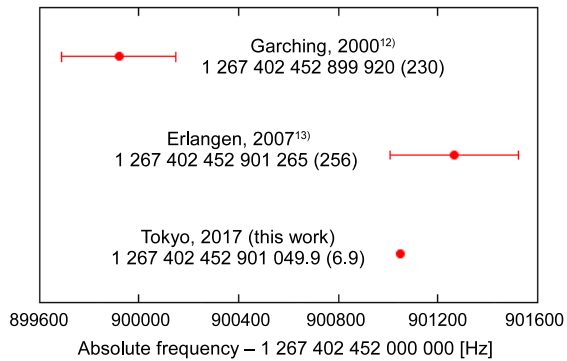}
\caption{Comparison of the measured frequency with previously reported values.}
\label{f-comparison}
\end{figure}

Figure \ref{f-comparison} represents the comparison of the measured frequency with reported values.
In the previous studies a methane-stabilized He-Ne laser or a commercial cesium beam clock both calibrated at remote sites was used as a frequency reference \cite{OL2000,OC2007}.
In contrast our measurement has been performed using two different frequency standards located in the same site, and, therefore, it presumably provides more reliable measurements.

\section{Conclusion}
In conclusion we have performed a measurement of the clock transition frequency of an In$^+$ with a fractional uncertainty of 5$\times$10$^{-15}$.
It is the first measurement of the In$^+$  transition by new approach using sympathetic cooling in a linear trap.
Our measurement determines the absolute clock transition frequency with the smallest uncertainty ever reported. 
The optical clock operation will be implemented by locking the clock laser to the transition, when the instability of the ion-chain arrangements is reduced by further studies.
The advantages of In$^+$ represented by the absence of the quadrupole shift and by the immunity to the BBR shift will be fully exploited in our future work including extension to the multi-ion optical clock.

\section*{Acknowledgments}
We thank H. Hachisu for his contribution in the operation of the Sr lattice clock, and R. Locke for his contribution in the early stage of the project. The stable optical cavity for 922nm laser was assembled by K. Etoh.
We are also grateful to M. Hosokawa and Y. Hanado for fruitful discussions. 
The project was achieved with the continuous technical help provided by H. Ishijima, M. Mizuno.
This work has been supported by JSPS and DAAD under the Japan-Germany Research Cooperative Program.


\end{document}